\documentstyle[12pt,a4]{article}
\begin{document}
\renewcommand{\baselinestretch}{1.5}

\title{The simplest {\em strange} three-body halo}
\author{A.~Cobis, A.~S.~Jensen and D.~V.~Fedorov\\
Institute of Physics and Astronomy, \\ Aarhus University, DK-8000
Aarhus C, Denmark \\} 
\date{\today}

\maketitle

\renewcommand{\baselinestretch}{1.0}

\begin{abstract} The recently developed method to solve the Faddeev
equations in coordinate space is used to study the weakly bound halo
nucleus $^{3}_{\Lambda}H (\Lambda + n + p)$. 
The long distances are treated carefully
to achieve convergence and high accuracy.  We use several sets of
two-body interactions which reproduce the deuteron properties and
provide the low-energy $\Lambda$-nucleon scattering data close to that
of two of Nijmegen potentials. We show that the details of the
potentials are unimportant unless the accuracy of the hypertriton
binding energy is required to be better than 50 keV.  We find that the
most significant parameter of the $\Lambda$-nucleon interaction, the
singlet s-wave scattering length, must be within 10\% of 1.85 fm, when
the $\Lambda$-separation energy from the deuteron is about 130 keV.
Other details of the $\Lambda$-nucleon interaction are less important
for the hypertriton structure. The scattering length and effective
range are computed for scattering of a $\Lambda$-particle on a
deuteron. The folding model reducing the three-body problem to a
two-body problem is investigated in this context and found to be
inadequate. \\

\end{abstract}

\renewcommand{\baselinestretch}{1.5}

\newpage

\section{Introduction}
The last decade has witnessed a rapid development following the
discovery of nuclear halos \cite{han95}. The most carefully studied
examples are $^{11}$Li and $^{6}$He \cite{zhu93}. The characteristic
features of halos are very small particle separation energies and very
large spatial extension \cite{alk96,tho94}. Simple few-body models are
used to describe these features and the details of the interactions
are not essential even in accurate computations of halo properties
\cite{fed94a,fed94b}.

It is interesting to extend the studies of "ordinary" nuclear halos to
systems with non-vanishing strangeness. The halo perspective should
then be kept and the general properties of halos extracted.  The
lightest nucleus with strangeness is the hypertriton
($^{3}_{\Lambda}H$). It consists of a neutron, a proton and a
$\Lambda$-particle. The system may roughly be described as a deuteron
and a $\Lambda$-particle bound in a state with the binding energy
$B_{\Lambda}$ = 0.13 $\pm$ 0.05 MeV \cite{jur73,dav91}. The deuteron
binding energy is $B_d=2.224575$ MeV and the total three-body binding
energy is then $B=2.35$ MeV.  The spin and the parity $J_d^{\pi} =
1^+$ and $J_{\Lambda}^{\pi} = \frac{1}{2}^{+}$, respectively of the
deuteron and the $\Lambda$-particle, are combined with the
corresponding quantum numbers of the relative state into the total
$J^{\pi} = \frac{1}{2}^{+}$ \cite{dal69}. If the deuteron remains
undisturbed within the hypertriton the total orbital angular
momentum can only take the values $L=0$ or 2. In any case, relative
s-states are probably the all-dominating components.

The hypertriton or the approximate $\Lambda$-deuteron two-body system
is very weakly bound and qualify as the simplest nuclear halo system
\cite{han95} with finite strangeness. The spatial extension of the
system is easily estimated by the root mean square radius $<r^2>^{1/2}
\approx \hbar/\sqrt{4\mu B_{\Lambda}} = 10.2$ fm, where the reduced
mass $\mu = m_{\Lambda}m_d/(m_{\Lambda}+m_d) = 0.760m_N$ is about 3/4
of the nucleon mass $m_Nc^2=939$ MeV. Thus the $\Lambda$-deuteron
radius is 5 times larger than the root mean square radius $R_d=1.971$
fm of the deuteron. The system appears to be very simple, but it is
close to the $\Lambda$-dripline and therefore very sensitive to the
strength of the $\Lambda$-nucleon interaction, which in this way
therefore is constrained by the measured hypertriton properties.

The hypertriton offers a mixture of simplicity, sensitivity and
fundamental interest in connection with the interactions between
nucleons and strange particles. This combination have attracted
considerable (theoretical) attention over the last three decades. The
interest is reflected in a number of investigations where different
techniques and models are used, e.g. variational models
\cite{kol88,che86}, the hyperspherical method \cite{ver79,cla85},
Faddeev calculations \cite{afn89,mig93} and simpler $\Lambda$-deuteron
two-body models \cite{con92}. Virtually all available interactions
have been used in these investigations including very sophisticated
microscopic interactions. Unfortunately the results also exhibit
considerable variation for example from the (observed) bound
three-body structure to the (wrong) unbound $\Lambda$-deuteron system
\cite{mig93}.

It is also very tempting to use the hypertriton to fine-tune the
$\Lambda$-nucleon two-body interaction. However, this certainly
requires very accurate knowledge of both the interactions between all
contributing partial waves of the subsystem and all contributing
couplings to other systems for example like the
$\Sigma$-nucleon-nucleon system. In addition the rather
delicate numerical three-body problem also must be mastered to a high
accuracy.

A convenient method to solve the quantum mechanical three-body problem
was recently formulated \cite{fed93a} and tested on various systems
for example on the numerically difficult problem of the occurrence of
Efimov states in nuclei \cite{fed93b}.  It is well-known
\cite{fed94a,fed94b} that one of the difficult three-body problems is
to give an accurate description of a system where one binary subsystem
has a loosely bound state and the third particle is weakly bound to
the other two particles. The characteristic size of the
$\Lambda$-deuteron system and the related small binding energy
requires a rather accurate treatment of the wave function to large
distances up to about 50 fm \cite{fed94b}. This is only barely
achieved in the best of the previous investigations.

Another interesting observation in this connection is that the large
extension of the system, which implies high sensitivity to the
two-body interactions, at the same time provides the simplification
that only low-energy properties of the $\Lambda$-nucleon interaction
is really needed in the description \cite{fed94a,fed94b}. These
properties may be expressed in terms of scattering lengths and
effective ranges of various relative states \cite{fed95}. Thus
extraction of other very detailed properties of the interaction from
the properties of the hypertriton might turn out to be very difficult.

The purpose of this paper is two-fold. First we want to develop,
refine and test the new method to solve the three-body problem
\cite{fed93a}. An accurate determination of the hypertriton structure,
including its binding energy, is a challenging problem due to the
bound subsystem and the weak binding energy \cite{fed94b}. The present
application is an investigation of a ``strange'' nuclear halo
system. In this connection it is worth noting that the previous
elaborate and complicated Faddeev calculations were obtained by using
different methods for example by working in momentum space. Results
from the present new method is desirable to get an independent
comparison. In addition to the bound state investigation we use the
occasion to apply the method in a strict three-body computation of
scattering of a $\Lambda$-particle on a deuteron which subsequently is compared to two-body scattering
calculations.  Secondly, we want
to investigate whether the low-energy properties of the
$\Lambda$-nucleon interaction suffice for the desired accurate
description of the hypertriton structure. In this connection we need
to study the effects of the nucleon-nucleon interaction, as reflected
in the deuteron properties, on the hypertriton structure.

This paper may be considered as part of a series discussing the general
properties of three-body halo systems. They all deal with weakly bound
and spatially extended three-body systems compared to the length and
energy of the two-body interactions. In the first of these papers
\cite{fed94a} we discussed Borromean systems where no subsystem is
bound \cite{zhu93}. In the second \cite{fed94b} we extended the
discussion to general three-body systems especially those where two
and three-body asymptotics are mixed. In the third \cite{fed95} we
discussed the effects of the finite spins of the particles. In the
fourth \cite{gar96} we discuss fragmentation interactions of halo
nuclei in the sudden approximation and especially the effects of final
state interactions.

In this paper dealing with a strange halo, we give in section 2 after
the introduction, a brief sketch of the method. In section 3 we
introduce and parametrize the all-dominating set of input parameters,
i.e. the two-body potentials. In section 4 we describe and discuss the
numerical results and in particular the strict three-body scattering
computations compared to the two-body approximation to this three-body
problem. Finally, in section 5 we offer a summary and the conclusions.

\section{Method}
The method is describe in previous publications, see for example
\cite{fed94a,fed94b,fed95}. Therefore we shall here only give a very
brief sketch, which should be sufficient to define the quantities of
interest and the notation used in the following sections. We consider
a system of three particles labeled by $i=1,2,3$ with corresponding
masses $m_{i}$, coordinates $\mbox{\bf r}_i$ and momenta $\mbox{\bf
p}_i$. They interact via the two-body potentials $V_{ij}$. We shall
use the Jacobi coordinates, which apart from mass factors, are defined
as the relative coordinates between two of the particles ($\mbox{\bf
x}$) and between their center of mass and the third particle ($\mbox{\bf
y}$).  The explicit definitions and the corresponding three sets of
hyperspherical coordinates ($\rho$,$\alpha$,$\Omega_x,\Omega_y$) can
be found in \cite{zhu93,fed94a,fed94b}.  Here $\rho~(=\sqrt{x^2+y^2})$
is the generalized radial coordinate and $\alpha$, in the interval
$[0,\pi/2]$, defines the relative size of $\mbox{\bf x}$ and $\mbox{\bf y}$,
where $\Omega_x$ and $\Omega_y$ are the angles describing the
directions of $\mbox{\bf x}$ and $\mbox{\bf y}$. The volume element is
given by $\rho^5{\rm d}\Omega{\rm d}\rho$ where ${\rm d}\Omega=\sin^2
\alpha \cos^2 \alpha {\rm d}\alpha {\rm d}\Omega_x {\rm d}\Omega_y$.

The total wavefunction $\Psi$ of the three-body system is given as a
sum of three components $\psi^{(i)}$ each expressed in terms of one of
the three different sets of Jacobi coordinates:
\begin{equation} \label{e4}
\Psi= \sum_{i=1}^{3}  \psi^{(i)}(\mbox{\bf x}_i,\mbox{\bf y}_i).
\end{equation}
These wavefunctions satisfy the three Faddeev equations
\begin{equation} \label{e5}
(T-E)\psi^{(i)} +V_{jk} (\psi^{(i)}+\psi^{(j)}+\psi^{(k)})=0,\;
\end{equation}
where $E$ is the total energy, $T$ is the kinetic energy operator and
$\{i,j,k\}$ is a cyclic permutation of $\{1,2,3\}$.  Each component
$\psi^{(i)}$ is now for each $\rho$ expanded in a complete set of
generalized angular functions $\Phi_n^{(i)}(\rho,\Omega_i)$.
\begin{equation} \label{e6}
	\psi^{(i)} = \frac {1}{\rho^{5/2}}
	\sum_n f_n(\rho)  \Phi_n^{(i)}(\rho ,\Omega_i)
	\; ,
\end{equation}
where the phase space factor $\rho^{-5/2}$ is extracted.  The angular
functions are now chosen for each $\rho$ as the eigenfunctions of the
angular part of the Faddeev equations
\begin{equation} \label{e7}
 {\hbar^2 \over 2m}\frac{1}{\rho^2}\hat\Lambda^2 \Phi_n^{(i)} +V_{jk}
(\Phi_n^{(i)}+\Phi_n^{(j)}+\Phi_n^{(k)})\equiv {\hbar^2 \over
2m}\frac{1}{\rho^2} \lambda_n (\rho) \Phi_n^{(i)} ,\;
\end{equation}
where we choose $m=m_N$ and $\hat\Lambda^2$ is the $\rho$-independent
part of the kinetic energy operator defined by
\begin{equation}\label{e8} 
      T \equiv T_{\rho}+{\hbar^2 \over 2m}\frac{1}{\rho^2}\hat\Lambda^2, \;
      T_{\rho}=-{\hbar^2 \over 2m}\left(
\rho^{-5/2}\frac{{\partial}^2}{{\partial}\rho^2}
\rho^{5/2}-\frac{1}{\rho^2} \frac{15}{4}\right) \; .
\end{equation}
Explicitly the generalized angular momentum operator $\hat \Lambda^2$
is given by
\begin{equation}  \label{e8a}
	\hat \Lambda^2=-{1 \over \sin\alpha\cos\alpha}
      {{\partial}^2\over {\partial} \alpha^2} \sin\alpha \cos\alpha 
	+{\hat {\mbox{\bf l}}_{x}^2 \over {\sin^2 \alpha}}
 	+{\hat {\mbox{\bf l}}_y^2 \over {\cos^2 \alpha}} -4 
\end{equation}
in terms of the orbital angular momentum operators ${\hat {\mbox{\bf
l}}}_{x}^2$ and ${\hat {\mbox{\bf l}}}_{y}^2$ related to the
$\mbox{\bf x}$ and $\mbox{\bf y}$ degrees of freedom.

The radial expansion coefficients $f_n(\rho)$ are solutions to the
coupled set of differential equations
\begin{equation} \label{e9}
   \left(-\frac{\rm d ^2}{\rm d \rho^2}
   -{2mE\over\hbar^2}+ \frac{1}{\rho^2}\left( \lambda_n(\rho) +
  \frac{15}{4}\right) \right)f_n
   +\sum_{n'}   \left(
   -2P_{nn'}{\rm d \over\rm d \rho}
   -Q_{nn'}
   \right)f_{n'}
   =0 , \;
\end{equation}
where the functions $P$ and $Q$ are defined by
\begin{equation}\label{e10}
   P_{nn'}(\rho)\equiv \sum_{i,j=1}^{3}
   \int d\Omega \Phi_n^{(i)\ast}(\rho,\Omega)
   {\partial\over\partial\rho}\Phi_{n'}^{(j)}(\rho,\Omega),\;
\end{equation}
\begin{equation}\label{e11}
   Q_{nn'}(\rho)\equiv \sum_{i,j=1}^{3}
   \int d\Omega \Phi_n^{(i)\ast}(\rho,\Omega) 
   {\partial^2\over\partial\rho^2}\Phi_{n'}^{(j)}(\rho,\Omega).\;
\end{equation}
The asymptotic large-distance behavior of the radial potentials can be
calculated in our case when one of the two-body subsystems has a bound
state with the binding energy $B_d$ and the corresponding
wavefunction $\phi(\mbox{\bf r})$. The result is obtained from the behavior
of the individual quantities 
\begin{equation}\label{e12}
 \lambda_1 \rightarrow -4 - \frac{2mB_d \rho^2}{\hbar^2} + 
 \langle \phi | 1 + 3 r {\partial \over \partial r} +
  r^2 {\partial^2\over\partial r^2} | \phi \rangle  \; ,
\end{equation}
\begin{equation}\label{e13}
 Q_{11} \rightarrow - \frac{1}{4\rho^2}  + 
 \frac{1}{\rho^2} \langle \phi | 1 + 3 r {\partial \over \partial r} +
  r^2 {\partial^2\over\partial r^2} | \phi \rangle    \; ,
\end{equation}
where the index 1 refers to the lowest $\lambda$-value.

Since the diagonal terms of $P$ vanish, i.e. $P_{nn}=0$, the diagonal
effective radial potential corresponding to the lowest $\lambda$ is
then given by
\begin{equation}\label{e14}
 \frac{1}{\rho^2} (\lambda_1 + \frac{15}{4}) - Q_{11} \rightarrow
 - \frac{2mB_d}{\hbar^2} + O(\rho^{-3})   \; .
\end{equation}
The diagonal effective potential for higher lying $\lambda$-values,
$(\lambda_n + 15/4)/\rho^2 - Q_{nn}$, instead approaches zero for $\rho
\rightarrow \infty$ at least as fast as $1/\rho^3$, see
\cite{fed94b,fed93a}.  The radial couplings arise from the
non-diagonal parts of $P$ and $Q$. They also approach zero for $\rho
\rightarrow \infty$ at least as fast as $1/\rho^3$.

\section{Two-body potentials}
The present three-body calculation explores how far it is possible to
understand the hypertriton as a three-body system from its constituent
particles and their mutual (phenomenological) interactions. The
low-energy properties of the interactions and the
interactions at distances much larger than the particle sizes are
decisive due to the relatively weak binding of the system. We assume
that the particles maintain their identity and their intrinsic structure
only enter via their spins and parities. The requirements for the
phenomenological two-body interactions are then first that they
accurately reproduce the low-energy properties of the two-body
subsystems. Secondly, the interactions must simultaneously be simple
enough to allow a flexible and accurate treatment of the three-body
problem. The latter requirement will be met by proper parametrization.

\subsection{The nucleon-nucleon interaction}
The main component of the relevant part of the nucleon-nucleon force
corresponds to the quantum numbers of the deuteron, i.e. the triplet s
and d-states with the total angular momentum and parity
$J^{\pi}=1^+$. However, in the three-body system also two-body
relative p-states are possible, since corresponding relative p-states
of the last particle can couple to give the proper total angular
momentum and parity. Higher orbital angular momenta are negligible.
The total isospin of the hypertriton is zero and the nucleon-nucleon
relative state then also must have vanishing isospin.

We parametrize the isospin zero part of the nucleon-nucleon
interaction with central, spin-spin, tensor and spin-orbit terms as
\begin{equation} \label{e21}
  V_{NN} = V_{c}(r) + V_{ss}(r) \mbox{\bf s}_{N1}
 \cdot\mbox{\bf s}_{N2} + V_{T}(r) 
 \hat{S}_{12} + V_{so}(r) \mbox{\bf l}_{NN}\cdot\mbox{\bf s}_{NN}
  \; ,
\end{equation} 
where $\mbox{\bf s}_{N1}$ and $\mbox{\bf s}_{N1}$ are the spins of the
nucleons, $\mbox{\bf s}_{NN} \equiv \mbox{\bf s}_{N1} + \mbox{\bf
s}_{N2}$, $\mbox{\bf l}_{NN}$ is the relative orbital angular momentum and
$\hat{S}_{12}$ is the usual tensor operator. It is sometimes
convenient to use a ``central'' potential including the spin-spin
part, i.e.
\begin{equation} \label{e22}
  V_{c}^{(2s+1)} = V_{c}(r) + V_{ss}(r) 
  \langle s| \mbox{\bf s}_{N1}\cdot\mbox{\bf s}_{N2} |s \rangle  \; ,
\end{equation} 
which then depends on the total spin s of the two-body system.  To
test the sensitivity of our results we want to vary the radial shapes
of the interactions while keeping the deuteron properties and the
low-energy scattering properties. We choose the radial shapes,
$V_{c}^{(1)}(r)$, $V_{c}^{(3)}(r)$, $V_{T}(r)$, $V_{so}(r)$, of each
of the terms to be either a gaussian $V$exp$({-r^2/b^2})$, an exponential 
$V$exp$({-r/b})$ or a Yukawa function $Vbr^{-1}$exp$({-r/b})$. 

The low-energy scattering data are available as scattering lengths $a$
and effective ranges $r_e$ of the different relative states. We also
have detailed experimental information about the deuteron
\cite{des95}, i.e. binding energy $B_{d} = 2.224575(9)$ MeV, root mean
square radius $R_d = <r^{2}>^{1/2} = 1.971(6)$ fm, admixture of the
d-state $P_{d} \sim 4\%-6\%$, the asymptotic ratio of the d to s-wave
component $\eta_d=0.0256(4)$ and electric quadrupole moment
$Q_d=0.2859(3)$ fm$^2$.

In the hypertriton wave function the deuteron component is essential
and reproducing the deuteron properties is therefore expected to be
more important than reproducing scattering data. As the s-wave by far
is the most dominating component we shall first explore this question
by using only the central part of the interaction and the
corresponding deuteron wave function without d-state admixture. We
then adjust the range and strength of the radial two-body potential to
reproduce the measured values of $B_{d}$ and $R_{d}$. We also varied
$R_{d}$ to see the influence of the deuteron size on the hypertriton
properties. The resulting parameters for these interactions labeled 1
and 2 can be found in the upper part of table \ref{tab1} for gaussian,
and exponential radial shapes. The many digits in the
parameters are necessary to obtain sufficient accuracy on the computed
quantities.

We now also include the d-wave admixture in the deuteron.  For
each radial shape we have then four ranges and four strengths to adjust
to the four singlet and triplet s-state low energy parameters \cite{dum83},
binding energy, root mean square radius, d-wave admixture, quadrupole
momentum and the asymptotic d- to s-wave ratio.  The parameters for
these interactions resulting from this consistent treatment are labeled
with an additional C and given in the middle part of table~\ref{tab1}.

In addition we constructed two more sets of potentials which provide
slightly different deuteron properties in order to check the
sensitivity of the resulting hypertriton structure to these
differences.  Those potentials are listed in the lower part of
table~\ref{tab1}.

The various properties of the deuteron for each of these
nucleon-nucleon interactions are shown in table \ref{tab2}. Even for
the simplest central interactions we obtain scattering lengths and
effective ranges for the triplet state in rather close agreement with
the measured values. The more complicated potentials from the middle
part of table \ref{tab1} also reproduce these scattering data rather
well. The quadrupole moment, the root mean square radius and the
asymptotic d to s-wave ratio are reproduced as well although with minor
deviations.  These quantities are nicely determined and some of the
remaining small deviations are in fact rather difficult to remove due
to strong model correlations between the different data.  The lower
part of table \ref{tab2} shows the properties of the additional sets of
potentials.  We shall use these slightly different properties of the
deuteron to study various specific correlation with the hypertriton
structure.

\subsection{The $\Lambda$-nucleon interaction}
The $\Lambda$-nucleon relative wave function within the hypertriton is
dominated by s-wave components. In the computations we shall also
allow p and d-state contributions, but we shall neglect the
contributions from the higher orbital angular momenta. The
parametrization of the interaction are again given by the form
\begin{equation} \label{e25}
 V^{(l)}_{\Lambda N} = V^{(l)}_{c}(r) + V_{ss}(r) 
 \mbox{\bf s}_{\Lambda}\cdot\mbox{\bf s}_{N} + V_{T}(r) 
 \hat{S}_{12} + V_{so}(r) \mbox{\bf
 l}_{\Lambda N}\cdot\mbox{\bf s}_{\Lambda N} \; ,
\end{equation} 
where $\mbox{\bf s}_{\Lambda}$, $\mbox{\bf s}_{N}$, and $\mbox{\bf
l}_{\Lambda N}$, respectively are the spin of the $\Lambda$-particle,
spin of the nucleon and the $\Lambda$-nucleon relative orbital angular
momentum. The total spin is $\mbox{\bf s}_{\Lambda N} \equiv \mbox{\bf
s}_{\Lambda} + \mbox{\bf s}_{N}$ and $l$ = 0,1. The central part may
have an orbital angular momentum dependence indicated by the
superscript $l$. This is to simplify simultaneous reproduction of all
the different p-wave scattering lengths. We shall also use a
parametrization without explicit $l$-dependence.

The low-energy properties of the interaction is not directly available
from scattering experiments. However, it is known that the
$\Lambda$-nucleon interaction does not bind the two-body system and
the binding energy of the hypertriton can therefore be used to
constrain models of this interaction. The role of the hypertriton and
the $\Lambda$-nucleon interaction for nuclear systems with finite
strangeness is in this way analogous to the role of the deuteron and
the corresponding nucleon-nucleon interaction for ordinary nuclei.

Several models are available for the $\Lambda$-nucleon interaction.  We
have selected two different models developed by the Nijmegen group.
They are one-boson exchange potentials obtained by using SU(3) symmetry
and simultaneous fits to all nucleon-nucleon and $\Lambda$-nucleon data
\cite{nag79,mae89,rij91}. We shall here focus on the models called {\em
Nijmegen F} and {\em Nijmegen SC} \cite{nag79,mae89}.  The computed
scattering lengths and effective ranges for s-states and the scattering
lengths for the p-states are shown in table \ref{tab3}. Using the
triplet p-state, this provides 7 constraints on our parametrization.
For convenience we do not use the singlet p-state in the fits, since it
is determined (as the s-waves) by the central and spin-spin parts of
the interaction and the corresponding contributions to the hypertriton
wave function are small. The low-energy parameters of the Nijmegen
models F and SC are reproduced by the interactions labeled 1, 3 and
5, 6 respectively.

As for the nucleon-nucleon case we determine strengths and ranges of
the potentials by adjusting to the model values of the s-wave
scattering lengths and effective ranges and triplet p-wave scattering
lengths given at the top of table~\ref{tab3}. The remaining degrees of
freedom are constrained by choosing the same ranges for some of the
terms. These choices are rather arbitrary, but they influence mostly
the small p-wave part of the $\Lambda$-nucleon relative wavefunction.
The resulting different sets of interaction parameters (label 1) are 
collected in table \ref{tab4}.

These interactions all overbind the hypertriton as we shall see later.
The binding energy is the decisive quantity determining the structure
of the halo and we must therefore adjust some of the interaction
parameters to reproduce the measured $B_{\Lambda}$. This can most
efficiently be achieved by changing the scattering length of the
dominating component in the wave function, i.e. the s-waves. For the
gaussian potential (G1) we then simply reduce the strengths
of the central s-wave potential by 10\% resulting in the interactions
G1r. For exponential and Yukawa shapes we then readjusted the
parameters of the potentials to reproduce these new slightly altered
low-energy scattering data.  The resulting parameters for these
interactions (Y1r and E1r) are also shown in table \ref{tab4}. 

As an alternative choice of the interaction parametrization we do not
allow now l-dependence in the central potential leaving 8 parameters
and 7 constraints. The eighth parameter is constraint by additional
fit to the $^3P_2$ effective range (labels 3 and 5). We use both the
Nijmegen potentials, although mainly Nijmegen SC, and reproduce the
scattering data with different radial shapes. Again the hypertriton
binding energy turns out to be too large. We reduce the strength of
(all partial waves of) the central potential by 10\%. The parameters
for the resulting interactions (labeled G5r, E5r and Y5r) are given in
table \ref{tab4}. To study the dependence on scattering lengths we varied 
the singlet and triplet s-wave parameters independently. The new sets 
are labeled "ra" and "rb" and the corresponding interaction parameters 
for gaussian, exponential and Yukawa shapes are also given in table \ref{tab4} 
(labeled G5ra, G5rb, E5ra, E5rb, Y5ra and Y5rb).

Finally, we have collected the scattering lengths and effective ranges
in table \ref{tab3} for all these $\Lambda$-nucleon interactions.

\section{Numerical results}
The method requires a priori specification of the quantum numbers of
the contributing components in all three Jacobi set of coordinates.  A
complete basis is obtained by including all possible values of these
quantum numbers consistent with the generalized Pauli principle,
parity and the rules for coupling angular momenta. The isospin
conservation requires that the two-nucleon system is in an isospin
zero state, i.e. a deuteron-like state.  Only very few of the lowest
orbital angular momenta contribute to the accuracy we need. However,
we shall include as many as necessary to obtain the required accuracy.

In the Jacobi coordinate system, where $\mbox{\bf x}$ refers to the
relative coordinate between the two nucleons, we need to include
orbital angular momenta $l_x=0,2$ with the corresponding spin values
$s_x=0$. The intermediate value $l_x=1$ (and the related $s_x=1$) is
not allowed in the two-body system due to parity conservation, but the
third particle can compensate by being in a relative p-state. In both
the other two Jacobi coordinate systems, where $\mbox{\bf x}$ refers to the
relative ${\Lambda}$-nucleon system, we expect again dominating
components with $(l_x,l_y)=(0,0)$ and admixture of components with
$(l_x,l_y)=(0,2),(2,0),(1,1)$. The d-wave components can be expected
to be similar to the d-wave admixture in the deuteron. The resulting
set of quantum numbers in our computations are given in table
\ref{tab5}.  A number of additional channels have also been included
to test the convergence. Their contributions always turned out to be
negligible in the present context.

\subsection{Convergence and accuracy}
The large extension or the weak binding of the system relative to a
free deuteron and ${\Lambda}$-particle demands an accurate
treatment. This means that the wave function must be calculated to
large distances and small components must be included both in the
angular Faddeev equations and in the coupled set of radial equations.
These convergence problems are essential as emphasized previously
\cite{ver79,cla85}.  The large distances are handled by analytical
extrapolation of the numerically obtained ${\lambda}$-values. This
improves significantly both accuracy and speed of the computation.

The separation energy $B_{\Lambda}$ is determined with a relative
accuracy of about $10^{-3}$ already by including the lowest three
values. This is roughly unchanged when the interactions result in both
somewhat weaker and stronger binding energy. This relative accuracy is
sufficient for our purpose and in the calculations we therefore always
included the three lowest ${\lambda}$-values in the radial set of
equations.

We first carry out a number of calculations using various combinations
of N-N and ${\Lambda}$-N interactions. They are parametrized to fit
some of the deuteron properties and the low-energy scattering data
obtained in the two Nijmegen models. The results are shown in table
\ref{tab6}.  Different radial shapes lead to differences in binding
energy of up to 340 keV out of the total energy of about 0.31 to 0.65
MeV. The N-N interactions which best fit the deuteron properties
(label~C) exhibit less binding and less deviations resulting from
different radial shapes. All the interactions considered in the upper 
part of table
\ref{tab6} produce a significant overbinding of the hypertriton. The
Nijmegen F model is always closest to the measured value. Since the
properties depend on the size and binding energy of the three-body
system, we adjusted the interactions to reproduce the hypertriton
binding energy. The resulting reduced potentials in table \ref{tab4}
are from now on used in the discussion.

The significance of different components in the wave functions can be
seen in the middle part of table \ref{tab6}. 
The contribution from the p-states amounts
to $\approx$ 10 keV compared to the value of $B_{\Lambda} \approx$ 60-157
keV. One exception is the Yukawa interaction in the ${\Lambda}$-N
interaction where the p-states contribute up to 40 keV.  These p-state
contributions are not allowed in the isolated deuteron wave function,
but in the hypertriton they are allowed in combination with the
negative parity components in the relative ${\Lambda}$-deuteron wave
function.

The contribution from the d-state in the ${\Lambda}$-nucleon channel
(13 and 14 in table \ref{tab5}) amounts to about 10-25 keV for most
interactions and to 55-70 keV for the nucleon-nucleon interactions
obtained by coupled channel analysis, where the tensor interaction
effectively is stronger. This d-wave contribution is mediated by the
d-wave in the deuteron and therefore only present when this admixture
is included. With d-wave admixture in the deuteron the binding energy
of the hypertriton is larger first due to the mentioned d-wave
contribution in the ${\Lambda}$-N channel and secondly due to the
deformation causing larger s-wave contribution.

The different radial shapes also marginally influence the
contributions. The wave function moves with decreasing binding energy
towards larger distances where the effective three-body potential is
determined by the scattering lengths of the two-body potentials. Thus
the detailed radial shapes should then be unimportant. However,
different radial shapes of the potentials always add a small
uncertainty.  In the present case of $B_{\Lambda} \approx$ 130 keV and
a 2.2 MeV bound binary subsystem this uncertainty arises almost
entirely from the ${\Lambda}$-N potential. The Yukawa potentials seems
to provide up to about 50 keV more binding than the exponentials which
in turn exceed the binding in gaussian potentials by about other 50 keV. 
The Yukawa potentials diverge at short distance which is
unphysical and rather extreme. The more reasonable finite potentials
like gaussians or exponentials lead to binding energies deviating by
less than 40 keV.  The deviations due to different radial shapes are
significantly diminished when the consistent analysis of the
scattering data are used to constrain the interactions. The resulting
variations in binding energy then amounts to less than 66 keV between
different shapes including Yukawa functions.

The sensitivity against variation of the radial shapes of the
potentials is investigated by introducing a repulsive core in the
central ${\Lambda}$-N potential. We simply use a linear combination of
two gaussians where the shortest range corresponds to $b=0.53$ MeV with
a strength of 550 MeV. The longer-range gaussian is then used in the
usual adjustment procedure where low-energy s-wave scattering data are
reproduced as described before. We only include central and spin-spin
terms. The resulting repulsive central potential at $r=0$ is then
$\sim$ 400 MeV. If the low-energy properties of all partial waves
should remain unchanged the l-dependent potential for each orbital
angular momentum must be individually adjusted. We only do this for s
and p-waves, see table \ref{tab4}. The effect of this repulsion is of
course the tendency for the wavefunction to decrease or vanish at short
distances between the two particles.

In an expansion on a basis, as we have used in the present work, this
requires more basis states with sufficient nodes at short distances.
Furthermore, the emphasis is somewhat shifted towards higher angular
momenta needed to describe the exclusion of the wavefunction from the
repulsive region. Both effects increase the difficulties in achieving
high accuracy and convergence in the numerical computations. The
typical results are a decrease of $B_{\Lambda}$ from 490 keV to 420
keV where we already included an energy gain of 22 keV by
adjusting the p-wave scattering to resemble the result for the case
without a repulsive core. For $B_{\Lambda}$ around 130 keV the total
decrease in binding energy is only 17 keV and the gain by p-wave
adjustment is reduced to about 17 keV, see table \ref{tab7}. 
The uncertainty related to a repulsive core is then less than the 
uncertainty due to different radial shapes.

Finally, we note that the mentioned total uncertainties of around 50
keV are extremely small compared to the central values of 10-100 MeV
of the two-body potentials holding the system together. Thus the
dependence on the details of the potentials is indeed remarkably
small. On the other hand the demand for high accuracy at some point
necessarily brings in some more details of the potentials. In the
present case this turns out to be at around 50 keV. Improved accuracy
can be achieved by an adjustment to the s-wave phase shifts over a
range of (low) energies. Then the choice of radial shape would be much
less important. The reason is simply that the decisive low-energy
properties deviate less over the contributing range of energies.

\subsection{Two-body interactions and the hypertriton structure}
The structure of the hypertriton described as a three-body system is
determined by the N-N and the ${\Lambda}$-N interactions. Since the
deuteron is a main component in this structure the most important
parts of the N-N interactions are those determining the deuteron
structure. The all-decisive property of the deuteron is the binding
energy and less importantly also the root mean square radius $R_d$ and
the d-wave admixture. We show $B_{\Lambda}$ in table \ref{tab6} for
different sizes and identical binding energies of the deuteron.  This
${\Lambda}$-separation energy decreases about 15 keV  when
$R_d$ increases from 1.904 fm to 1.971 fm. The variation of the radial
shape of the N-N interaction only causes insignificant changes when both
$B_d$ and $R_d$ are fixed. The inherent uncertainty in these
calculations is therefore only a few keV, since the N-N interaction and
the deuteron properties both are very well known.

The sensitivity to the ${\Lambda}$-N interaction is higher because the
system is close to the threshold for binding the
${\Lambda}$-particle. Then only a little change of the potential will
change the size and binding dramatically. Since the interaction is
less known it is of considerable interest to determine the decisive
degrees of freedom. This will provide information about which
constraints the hypertriton properties can place on the interaction.
The numerical examples in table \ref{tab6} show that different
interactions with the same radial shape and the same s and p
low-energy scattering data lead to the same hypertriton
properties. Different radial shapes (i.e., gaussian and exponential) 
and identical scattering data
produce a variation of $B_{\Lambda}$ of about 50 keV. 

The same radial shape and a 10\% bigger singlet s-wave scattering 
length (ra) increases $B_{\Lambda}$
by about 50-100 keV, respectively for gaussian and exponential shapes. 
A 10\% decrease of the triplet s-wave scattering length (rb) decreases 
the $B_{\Lambda}$ only by about 15-30 keV, for both gaussian and exponential 
shapes of the interaction. At least the dependence on the
singlet scattering length is significant in the present context.
Thus, accurate knowledge of the binding energy constrains the s-wave
scattering lengths and especially the singlet s-wave. The present
accuracy on $B_{\Lambda}$ allows the claim of about 10\% deviation
from the value of the singlet s-wave scattering length for the 
interaction Fr. The original two potentials, Nijmegen F and Nijmegen SC, 
certainly both overbind the hypertriton. Their singlet s-wave scattering 
lengths are both too large.

The separation energy $B_{\Lambda}$ has so far been in focus, but the
root mean square radius is another important and strongly correlated
quantity. It is shown in table \ref{tab6} for the
different interactions. In these tables we also show the mean distance
of the ${\Lambda}$-particle from the center of mass of the deuteron
under the assumption that the deuteron is unperturbed in the
hypertriton. The precise definition is given by
\begin{equation} \label{e35}
<r^{2}_{\Lambda d}> = \frac{m_N(m_{\Lambda}+m_d)}{m_{\Lambda}m_d}  
 (<\rho^2> - 2R_d^2) \; .
\end{equation}
This distance is much larger than the size of the deuteron and the
$\Lambda$-particle is therefore on average far outside the deuteron.
It is easy to verify that the product $B_{\Lambda}\cdot<r^{2}_{\Lambda d}>$
remains approximately constant. The reason for this two-body halo
dependence is the small separation energy from a ``particle''
asymptotically approaching the deuteron.

We are now in a position to show the realistic structure of the
hypertriton. First, we recall that the angular eigenvalues $\lambda$
are closely related to the diagonal part of the effective radial
potentials, $(\lambda_n +15/4)/\rho^2 - Q_{nn}$. The $\lambda$-spectrum
is therefore decisive for the spatial structure of the system. An
example including the five lowest values is shown in fig.1 for the GC1
and the G5r interactions respectively for N-N and $\Lambda$-N. The
behavior is typical and other realistic potentials like the
combination GC1 and Y5r or GC1 and E5r would differ by about the 
thickness of the lines already after $\sim$ 3 fm (see fig.1a).

At $\rho=0$, we have the hyperspherical spectrum $\lambda=K(K+4)$,
where K is a non-negative even integer. The odd numbers are excluded here
for parity reasons.  The lowest $\lambda$ starts at zero, has a flat
minimum and then bends over and diverges at large distances
parabolically as $\lambda = -2B_dm_N\rho^2/\hbar^2$. The other
$\lambda$-functions reestablish the hyperspherical spectrum
asymptotically at large $\rho$. Therefore they all approach finite
constants at large $\rho$. In particular the lowest level originating
from $\lambda=12$ is not far from the lowest $\lambda$-value at small
distance close to the attractive pocket. This level, which approaches zero
for $\rho \rightarrow \infty$, is responsible for the amount of relative
p-states in the subsystems.

The radial wave functions corresponding to the three lowest
$\lambda$-values are shown in fig.2. By far the most dominating
component is that of the lowest $\lambda$, which at large distances
describes a $\Lambda$-particle bound to a deuteron in the ground
state. The main part of the probability is found between 2 fm and 10
fm. The other components are very small, but they are essential to
achieve the accuracy we want. 
 
A more visual impression of the hypertriton structure is obtained from
fig.3, where we plot the probability distribution in the plane defined
by the three particles. We use the system of coordinates where the
center of mass is in the middle of the figure, the largest principal
moment of inertia is horizontal and the $\Lambda$-particle is in the
right half. For simplicity, only s-waves are included in the
computation and we used G2 N-N and G6r $\Lambda$-N interactions (see
tables \ref{tab1} and \ref{tab4}). The probability distribution
reaches its maximum at the point where the hyperradius is equal to
6.20~fm, the hyperangle is equal to 12.27 and the angle between ${\bf
x}$ and ${\bf y}$ is $\pi/2$. At this point the distance between the
nucleons is 1.86 fm and the distance between their center of mass and
the $\Lambda$-particle is 7.65 fm.

\subsection{Scattering of a $\Lambda$-particle on a deuteron}
The properties of the hypertriton are also accessible through
information obtained by scattering of $\Lambda$-particles on
deuteron. The $\Lambda$-N interaction could be further constrained
by such scattering measurements. If such information should become
available and subsequently utilized the corresponding three-body
scattering problem must be solved. 
Our method, sketched in section 2, allows computation of bound states as
well as scattering problems. Both types of computations are three-body
calculations and in particular the scattering problem does not have to
rely on two-body or any other approximation. 

The procedure is analogous to the bound state problem. The angular
eigenvalues $\lambda$ are first computed for each value of $\rho$.
Then the coupled set of radial equations are solved with the proper
boundary conditions, i.e. exponentially decreasing radial functions
$f_n$ except for the function corresponding to the lowest
$\lambda$-value, which asymptotically describes the deuteron in its
ground state. This function is at large distances turning into a phase
shifted sine-function in the variable $ky$ or equivalently $k\rho$.
When the energy of the $\Lambda$-particle relative to the deuteron
approaches zero we may in the usual way derive the scattering length
$a_{\rho}$ and effective range $r_{\rho}$ from these phase
shifts. 

These quantities are then related to the $\rho$ degree of freedom as
indicated by the notation. They must then be transformed to the
coordinates describing the relative distance between the deuteron and
the $\Lambda$-particle. The resulting scattering length and effective
range are then given by
\begin{equation} \label{e37}
  a_{\Lambda-d}= a_{\rho}\sqrt{\frac{m_N(m_{\Lambda}+m_d)}{m_{\Lambda}m_d}} 
  \;   \; ,  \; \;  
  r_{\Lambda-d} = r_{\rho}\sqrt{\frac{m_N(m_{\Lambda}+m_d)}{m_{\Lambda}m_d}} 
     \; .
\end{equation}

The hypertriton has angular momentum $\frac{1}{2}$, which means that
the deuteron spin of 1 couples to the spin $\frac{1}{2}$ of the
$\Lambda$-particle. Thus the relative $\Lambda$-deuteron state or
equivalently the scattering channel has spin $\frac{1}{2}$.

In the present case we aimed directly at computation of the scattering
length and effective range by using zero energy from the
beginning. Then the lowest radial wave function is a straight line
outside the short-range potential. The intersection with the
$\rho$-axis gives the scattering length $a_{\rho}$($^{2}S$). The
effective range $r_{\Lambda-d}$($^{2}S$) is obtained by direct
integration of the expression
\begin{equation} \label{e39}
  r_{\Lambda-d}(^{2}S) = 2 \sqrt{\frac{m_N(m_{\Lambda}+m_d)}{m_{\Lambda}m_d}} 
 \int_0^{\rho_{max}}\left(\tilde{f}_0^2 -  {f}_0^2\right) d\rho \; ,
\end{equation}
where $f_0$, corresponding to the lowest $\lambda$-value, is the
radial wave function normalized to unity at the origin. The related
wave function $\tilde{f}_0$ is the radial solution without any
interaction. It is identical to $f_0$ for distances larger than
$\rho_{max}$ or equivalently outside the short-range potential.

Using three $\lambda$-values as for the bound state calculation, we
obtain a scattering length accurate up to about 1\% and an effective
range accurate up to about 3\%. Both quantities are given in the
table \ref{tab6} along with other properties of the
hypertriton for the various cases considered.  The large value of
$a_{\Lambda-d}$($^{2}S$) is a natural consequence of the small
$\Lambda$-separation energy $B_\Lambda$. The connection is
approximately given by
\begin{equation} \label{e40}
B_\Lambda = \frac{\hbar^2}{2\mu a_{\Lambda-d}^2(^{2}S)}
  \frac{1}{1+r_{\Lambda-d}(^{2}S)/a_{\Lambda-d}(^{2}S)}  \; ,
\label{ascat}
\end{equation}
which numerically is fulfilled. The effective ranges are
in all cases about 4.0 fm $\pm$ 0.8 fm, i.e. as usual of the same
order as the range of the effective potentials.

\subsection{Two-body approximations to the three-body problem }
Reduction of a three-body problem to an effective two-body problem
might be a computational advantage. It is tempting to try such a
reduction for the hypertriton which approximately can be considered to be a
$\Lambda$-particle and a deuteron. This case appears to be very well
suited for a two-body description as also attempted previously
\cite{con92}.  The first step of the procedure is obviously to
construct the $\Lambda$-deuteron effective potential by folding the
deuteron wave function with the $\Lambda$-N interaction. The remaining
steps are then of two-body nature and therefore computationally much
simpler.

The interaction between the $\Lambda$-particle and the deuteron is
given by
\begin{equation} \label{e41}
  V_{\Lambda d}(\mbox{\bf r}_{\Lambda d}) = \int d\mbox{\bf r}^3_d  
 \psi_d^*(\mbox{\bf r}_d) \left(
 V_{\Lambda N}(\mbox{\bf r}_{\Lambda}-\mbox{\bf r}_{N1}) + 
 V_{\Lambda N}(\mbox{\bf r}_{\Lambda}-\mbox{\bf r}_{N2})
   \right)   \psi_d(\mbox{\bf r}_d)   \; ,
\label{twobody}
\end{equation}
where $\psi_d$ is the deuteron wave function obtained by solving the
Schr\"{o}dinger equation with the nucleon-nucleon potential. The
solution $\psi_{\Lambda d}$ to the Schr\"{o}dinger equation with this
effective two-body potential then gives an approximation to the
original three-body problem, i.e. the product wave function
$\psi_{\Lambda d}(\mbox{\bf r}_{\Lambda d}) \psi_d(\mbox{\bf r}_d)$.

The large-distance asymptotic behavior of the correct three-body wave
function describes a $\Lambda$-particle bound by $B_{\Lambda}$ to a
deuteron in the ground state. The relative coordinate $\mbox{\bf
r}_{\Lambda d} \equiv \mbox{\bf y} \sqrt{m_N
(m_{\Lambda}+m_d)/(m_{\Lambda} m_d) }$ is then, apart from the mass
factor, identical to $\rho (\approx y)$ at large distance. After
correcting for this scale factor on the coordinate, the radial
potential in the three-body computation should therefore be directly
comparable to the two-body potential.

Fig.4 shows this comparison between the potential corresponding to the
lowest $\lambda$-value and the folding result where for simplicity only
s-waves in the ${\Lambda}$-nucleon channel were included. The
differences at short distance obviously arise from the spatial
extension of the deuteron causing the coordinates to differ. The
differences at large distance are also significant. The folding
potential necessarily falls off exponentially in accordance with the
exponential behavior of the deuteron wave function and the short-range
${\Lambda}$-nucleon potential. On the other hand the three-body
potential decreases faster than $\rho^{-3}$, but presumably still much
slower than exponentially. The reason is that the three-body potential
also must account for polarization effects where the deuteron wave
function in the hypertriton is distorted away from its ground state
shape. Still asymptotically, but not at intermediate and short
distances, the potentials both correspond to the deuteron in its ground
state.

The hypertriton is in this two-body approximation always unbound, but
scattering in the two-body approximation can still be studied and
compared with the full three-body results.  With a binding energy
$B_{\Lambda}\approx 130$ keV the two-body approximation leads to the
scattering length and effective range around 4.8 fm and 3.7
fm respectively.  Through eq. (\ref{ascat}) this corresponds to a
two-body virtual state at around 680 keV, i.e. about 800 keV above the
correct energy value. Therefore we conclude that the folding model
cannot be used to estimate the properties of loosely bound systems like
the hypertriton.  The three-body calculations are then necessary to
obtain detailed information.

On the other hand an effective two-body model might sometimes be
useful in qualitative understanding and in preliminary estimates. We
can perhaps get some insight by specifying that the folding
approximation in eq. (\ref{twobody}) basically assumes a specific form for the
total three-body wave function. The resulting energy is therefore
only an upper bound and the hypertriton comes out unbound by about
500 keV. Clearly the potential must be more attractive than the folding
potential if the separation should be reproduced. 

Increasing the depth and/or the range parameter until the measured
binding energy is reached then provide a two-body model for the
hypertriton. The distance between $\Lambda$ and the deuteron, measured
for example as the root mean square radius, is then in this model too
small due to the too attractive potential needed to compensate for the
inflexible assumption of the wave function. The smaller the
$\Lambda$-deuteron separation energy the more extended the system and
the better is the two-body model. However, processes involving the
intrinsic structure with one neutron and one proton can not be
described.

An analogous example is the popular $^{11}$Li described as a three-body
system consisting of $^{9}$Li surrounded by two neutrons. Again the
intrinsic structure can not be described in the model. The distances
between these three particles are of comparable size and detailed and
accurate information therefore require three-body
calculations. Contrary to the hypertriton with one bound binary
subsystem, the Borromean system $^{11}$Li would be progressively worse
described for decreasing two-neutron separation energy.

\section{Summary and conclusion}
The general properties of nuclear three-body halo systems have been
discussed in a recent series of papers. The characteristic features of
these bound states are the weak binding and the large spatial
extension. The hypertriton certainly qualifies as a halo system due to
the relatively weak interactions between the $\Lambda$-particle and
the nucleons. Thus strangeness is involved and this system is
potentially able to provide information of a very different type from
that of the usual halo nuclei.

To relate any available experimental information with the basic
quantities determining the hypertriton structure, it is necessary to
perform accurate and reliable computations. We solve the coordinate
space Faddeev equations by use of a recently developed method where
the angular part first is obtained for any given average distance of
the particles. The large distance behavior is known analytically and
used to improve accuracy and speed of the calculations. The coupled
radial equations are solved numerically afterwards. The essential
large-distance behavior is in this way carefully treated as required
by the small binding and the related large extension.

We first sketch the method and then we parametrize the crucial input
for the calculations, i.e. the two-body interactions. We use
potentials with central, spin-spin, spin-orbit and tensor terms.  To
study the sensitivity we parametrize each of these with gaussian,
exponential and Yukawa radial shapes and compare the results.  For the
nucleon-nucleon part we adjust the parameters to constraints obtained
from the measured deuteron properties and the s-wave scattering
lengths and effective ranges. The $\Lambda$-nucleon interaction is
much less known. We use the low-energy parameters provided by two
potentials from the Nijmegen group obtained by SU(3) symmetry and fits
to all available nucleon-nucleon and $\Lambda$-nucleon data. We adjust
the parameters to reproduce the corresponding s-wave scattering
lengths and effective ranges and the p-wave scattering lengths. We
compare results where d-wave coupling is neglected in the two-body
scattering analysis used for the parameter extraction.
 
The dependence of the hypertriton properties on the various choices
are investigated in some detail. First the convergence and accuracy of
the computations are investigated. Three radial equations suffice to
an accuracy of 1 part in 1000. The sensitivity to the nucleon-nucleon
interaction is rather low provided the deuteron properties are
maintained. Both $\Lambda$-nucleon potential models overbind the
hypertriton by up to half an MeV. Since the binding energy is the
crucial quantity determining the size of the system, we reduced the
attractive strength of the $\Lambda$-nucleon central potential by
10\%. The hypertriton energy is then roughly correct.  For a larger
binding energy the $\Lambda$-particle is on average closer to the
deuteron and the hypertriton structure is more sensitive to the
details of the two-body potentials. Most effects would then be
amplified.  The sensitivities are more realistic when the measured
binding energy is used as starting point in the investigations.

The p-wave contribution to the binding energy is small but visible. A
typical value is 10 keV. It adds a negative parity component to the
relative neutron-proton wave function. In the three-body system this
is compensated by another p-wave in the relative $\Lambda$-deuteron
wave function. The contribution to the binding energy from the d-wave
in the $\Lambda$-nucleon relative state is also small but visible with
a typical value of about 20 keV.

The most essential part of the $\Lambda$-nucleon interaction
corresponds to the singlet s-wave. A 10\% change of this scattering
length changes the binding energy by about 70\% around the
$\Lambda$-separation energy $B_{\Lambda} \approx$ 130 keV. Changing
towards a weaker potential easily then results in an unbound
system. For comparison, a change of 10\% of the triplet s-wave
scattering length changes $B_{\Lambda}$ by about 15 keV. The other
scattering lengths are less important in this connection. To obtain
the correct binding energy $B_{\Lambda}$ we must therefore have a
singlet scattering length within 10\% from the value 1.85 fm.

The uncertainty arising from the different choices of radial shapes of
the $\Lambda$-nucleon potential amounts to 50 keV. A repulsive core
of up to 550 MeV introduces an additional inaccuracy of about 15 keV
for a $\Lambda$-separation energy of 130 keV. These uncertainties are
significant when compared to the $\Lambda$-separation energy of about
130 keV, but extremely small compared to the strengths of the two-body
potentials. This uncertainty is in fact also very small compared to
2.35 MeV which is the total hypertriton binding energy including the
deuteron part. Thus the chosen low-energy scattering data determines
the hypertriton structure and its binding energy to an accuracy of
about 50 keV.

The origin of the differences due to the radial shapes evidently must
arise from finer details of the potentials. The differences would be
reduced by adjusting to phase shifts in an energy range extending
beyond that reproduced by the scattering lengths and the effective
ranges. This corresponds to a better description of the higher-energy
or off-shell behavior. The extreme low-energy scattering properties we
used in the fits can only predict structures to the accuracy 50 keV
specified here. The lower the binding energy the better can the
properties be predicted by the extreme low-energy scattering behavior.

The three-body scattering problem, where a $\Lambda$-particle is
scattered on a deuteron, must also contain information about the
hypertriton structure or the two-body interactions describing this
three-body system. Using our method we carried out strict three-body
calculations of scattering length and effective range. 
The scattering length is closely related to
the size of $B_{\Lambda}$ and the effective range is of the order of
the range of the effective three-body radial potential.

We also tested the two-body approximation obtained by folding the
deuteron wave function and the ${\Lambda}$-nucleon interaction. We
computed binding energies and scattering properties. The approximation
is inadequate for the details and the accuracy required in the present
investigation.

In conclusion, the hypertriton is a challenging system where a careful
treatment is necessary to connect measured values with the basic
interaction involving strangeness. The halo structure makes the large
distances important and the numerical work difficult. The details of
the interactions enter rather weakly in the hypertriton structure
which predominantly is determined by a few of the scattering
lengths. Thus further constraints on the interaction would certainly
require considerably higher accuracy both in measurements and in
calculations.

{\bf Acknowledgments} One of us (A.C.) acknowledges support
from the European Union through the Human Capital and Mobility program
contract nr. ERBCHBGCT930320. The numerous enchanting discussions with
E.~Garrido are also kindly acknowledged.

\newpage

\noindent{\Large\bf{Figure Captions}}
\begin{list}{}{\setlength{\leftmargin}{18mm}
\setlength{\labelwidth}{16mm}
\setlength{\labelsep}{2mm}}

\item[Figure 1\hfill] The angular eigenvalue spectrum $\lambda$ as
function of $\rho$ corresponding to the interactions GC1 and G5r,
respectively for the N-N and $\Lambda$-N interactions. The five lowest
$\lambda$-values are exhibited. a) Short-distance behavior of the four
lowest $\lambda$-values for different $\Lambda$-N interaction radial
shapes.

\item[Figure 2\hfill] The components of the wave function for the
lowest three $\lambda$-values corresponding to the calculation in
figure 1.

\item[Figure 3\hfill] The probability distribution for the hypertriton
in the plane drawn through the three particles. The system of
coordinates is defined by the following constraints: i) the center of
mass is in the middle of the figure; ii) the principal moment of inertia
is along horizontal axes; iii) the $\Lambda$-particle is in the right
half of the figure. Only s-waves are included in the computations. 
The G2 and G6r interactions, respectively for the N-N and the 
$\Lambda$-N subsystems, are used (see tables \ref{tab1} 
and \ref{tab4}). Regions with higher z values are darker.

\item[Figure 4\hfill] The folding potential $V_{\Lambda d}$ (dot-dashed
curve) compared to the effective three-body radial potential $V_{eff}$
(solid curve) for the interactions GC1 and G6r. The coordinate used is
$\rho$ and the mass corrected distance between the $\Lambda$-particle
and the deuteron  $y=r_{\Lambda d}\sqrt{(m_{\Lambda} m_d)/m_N
(m_{\Lambda}+m_d)}$.

\end{list}

\newpage

\begin{table}[t]
\renewcommand{\baselinestretch}{0.9}
\caption{Parameters of the nucleon-nucleon (N-N)
interaction. Strengths $V$ in MeV and ranges $r$ in fm of the singlet
and triplet central (or central (c) and spin-spin (ss)), tensor (t)
and spin-orbit (so) parts are listed. The radial shapes are gaussian (G), 
exponential (E) and Yukawa (Y), respectively. The upper part of the table 
refers to a deuteron in the $^{3}S_{1}$ state; the lower part refers to a 
deuteron in the coupled $^{3}S_{1}$-$^{3}D_{1}$ two-nucleon state.}
\renewcommand{\baselinestretch}{1.5}
\vspace{5mm} 
\begin{center}
\begin{tabular}{c|cccccccc} \hline
  N-N & $V_{c}^{(3)}$ & $r_{c}^{(3)}$ & $V_{c}^{(1)}$ & $r_{1}^{(1)}$ & $V_{t}$ & $r_{t}$ & $V_{so}$ & $r_{so}$ \\ \hline
 G1 & -79.852 & 1.3971 & 0 & - & 0 & - & 0 & - \\
 E1 & -221.048 & 0.6129 & 0 & - & 0 & - & 0 & - \\
 G2 & -61.263  & 1.6383 & 0 & - & 0 & - & 0 & - \\
 E2 & -164.994 & 0.7291 & 0 & - & 0 & - & 0 & - \\ 
\hline
 GC1 & -49.768  & 1.658  & -31.051  & 1.8164 & -8.636  & 2.1786 & 110.347 & 1.4838 \\ 
 EC1 & -127.489 & 0.7997 & -105.072 & 0.7270 & -27.216 & 0.6552 & 322.983 & 0.7309 \\ \hline
  N-N & $V_{c}$ & $r_{c}$ & $V_{ss}$ & $r_{ss}$ & $V_{t}$ & $r_{t}$ & $V_{so}$ & $r_{so}$ \\ \hline
 G3 & -55.714 & 1.6184 & -72.250 & 0.9712 & -14.33   &  0.59 &  89.0 & 1.9 \\
 Y3 & -30.408 & 1.8013 & -6.879 & 2.6214 & -1.951 & 0.9703 &  68.6 & 1.9701 \\
 \end{tabular}\\ 
\label{tab1}
\end{center}
\end{table}

\begin{table}[b]
\renewcommand{\baselinestretch}{0.9}
\caption{Properties of the deuteron for the various interactions in
table \protect\ref{tab1}. The columns contain the root mean square 
radius $<\!r^2\!>^{1/2}$, the
$d$-state probability $P_{d}$, the electric quadrupole momentum $Q_d$,
the asymptotic d to s-wave ratio $\eta_d$, and the s-wave scattering
lengths $a$ and effective ranges $r_e$. The deuteron binding energy is
$B_{d} = 2.224575$ MeV. The first row shows the experimental values.
Note that in our sign convention a negative scattering length
corresponds to a bound state.}

\renewcommand{\baselinestretch}{1.5} 
\vspace{5mm} 
\hspace*{-1.5cm}
\begin{tabular}{c|cccccccc} \hline N-N & $<\!r^2\!>^{1/2}$ & $P_{d}$ &
$Q_d$ & $\eta_d$ & $a(^{3}S_{1})$ & $r_e(^{3}S_{1})$ & $a(^{1}S_{0})$ &
$r_e(^{1}S_{0})$ \\ & (fm) & (\%) & (fm$^{2}$) & & (fm) & (fm) & (fm) &
(fm) \\ \hline exp\cite{des95,dum83} & 1.971(6) & 4-6 & 0.2859(3) &
0.0256(4) & -5.4194(20) & 1.759(5) & 23.748(10) & 2.75(5) \\ \hline 
G1  & 1.904 & 0 & 0 & 0 & -5.324 & 1.620 & - & - \\ 
E1  & 1.904 & 0 & 0 & 0 & -5.330 & 1.598 & - & - \\ 
G2  & 1.971 & 0 & 0 & 0 & -5.485 & 1.836 & - & - \\ 
E2  & 1.971 & 0 & 0 & 0 & -5.485 & 1.799 & - & - \\ 
\hline 
GC1 & 1.967 & 5.66 & 0.273 & 0.0233 & -5.419 & 1.93 & 23.748 & 2.75 \\ 
EC1 & 1.971 & 5.63 & 0.271 & 0.0227 & -5.394 & 1.87 & 23.748 & 2.75 \\ \hline 
G3  & 1.904 & 4.07 & 0.24 & 0.0266 & -4.615 & 1.425 & 23.748 & 2.75 \\ 
Y3  & 2.039 & 4.07 & 0.35 & 0.0371 & -4.397 & 1.563 & 23.748 & 2.75 \\ 
\end{tabular} \\
\label{tab2} 
\end{table}

\begin{table}[p]

\renewcommand{\baselinestretch}{0.9}

\caption{Scattering lengths and effective ranges in fm for the two
selected $\Lambda$-nucleon ($\Lambda$-N) models
\protect\cite{nag79,mae89}. The rows labeled Fr and SCr represent the
values obtained by a 10\% reduction of the s-wave central strength of
the gaussian interactions, which otherwise reproduces the values for
these quantities obtained from the two models. The rows labeled ra and
rb represent variations where either the singlet S or the triplet S
scattering lengths are changed from the values of Fr and SCr by 10\%,
respectively up or down. Finally, the two last rows represent the 
the s-wave low-energy data of the models SC and SCr respectively,
reproduced by using gaussian interactions with only central and spin-spin 
terms, with (label cp) and without repulsive core, see table \protect\ref{tab4}.}

\renewcommand{\baselinestretch}{1.5}

\begin{center}
 \begin{tabular}{lcccccccc} 
\hline
$\Lambda$-N &$a(^{1}S_{0})$ & $r_e(^{1}S_{0})$ & $a(^{3}S_{1})$ & $r_e(^{3}S_{1})$ & 
$a(^{1}P_{1})$ & $a(^{3}P_{0})$ & $a(^{3}P_{1})$ & $a(^{3}P_{2})$  \\ 
\hline \hline
F\cite{nag79}  & 2.29 & 3.17 & 1.88 & 3.36 & 0.047 & -0.114 & 0.02  & -0.188 \\
Fr   & 1.85 & 3.41 & 1.52 & 3.64 & 0.047 & -0.114 & 0.02  & -0.188 \\ 
\hline
G1   & 2.30 & 3.17 & 1.90 & 3.35 & 0.111 & -0.114 & 0.020 & -0.118 \\
Y1   & 2.31 & 3.21 & 1.90 & 3.40 & 0.107 & -0.099 & 0.016 & -0.103 \\
G3   & 2.29 & 3.16 & 1.88 & 3.36 & 0.421 & -0.114 & 0.020 & -0.686 \\ 
G1r  & 1.86 & 3.40 & 1.54 & 3.63 & 0.111 & -0.114 & 0.020 & -0.118 \\
E1r  & 1.86 & 3.47 & 1.53 & 3.69 & 0.150 & -0.108 & 0.018 & -0.178 \\
Y1r  & 1.89 & 3.68 & 1.55 & 3.85 & 0.115 & -0.097 & 0.019 & -0.100 \\
\hline \hline
SC\cite{mae89} & 2.78 & 2.88 & 1.41 & 3.11 & 0.062 & -0.096 & 0.061 & -0.20 \\ 
SCr  & 2.27 & 3.10 & 1.15 & 3.36 & 0.41  & -0.068 & 0.086 & -0.42 \\ 
\hline
G5   & 2.78 & 2.90 & 1.41 & 3.11 & 0.437 & -0.096 & 0.061 & -0.448 \\ 
E5   & 2.78 & 2.88 & 1.41 & 3.57 & 0.442 & -0.128 & 0.061 & -0.551 \\ 
Y5   & 2.78 & 2.88 & 1.41 & 3.42 & 0.432 & -0.059 & 0.062 & -0.525 \\ 
G5r  & 2.27 & 3.10 & 1.15 & 3.36 & 0.405 & -0.068 & 0.086 & -0.415 \\ 
E5r  & 2.27 & 3.10 & 1.15 & 3.54 & 0.407 & -0.075 & 0.085 & -0.451 \\ 
Y5r  & 2.27 & 3.10 & 1.16 & 3.85 & 0.397 & -0.066 & 0.085 & -0.494 \\ 
G5ra & 2.50 & 3.11 & 1.15 & 3.36 & 0.451 & -0.067 & 0.085 & -0.416 \\ 
E5ra & 2.49 & 3.10 & 1.15 & 3.48 & 0.446 & -0.072 & 0.086 & -0.442 \\ 
Y5ra & 2.49 & 3.10 & 1.15 & 3.57 & 0.428 & -0.065 & 0.086 & -0.457 \\ 
G5rb & 2.27 & 3.11 & 1.04 & 3.36 & 0.408 & -0.064 & 0.087 & -0.366 \\ 
E5rb & 2.27 & 3.10 & 1.04 & 3.64 & 0.407 & -0.067 & 0.086 & -0.413 \\ 
Y5rb & 2.27 & 3.13 & 1.05 & 3.87 & 0.399 & -0.053 & 0.087 & -0.450 \\ 
\hline
G6, G6cp & 2.78 & 2.88 & 1.41 & 3.11 & 0.434 & -0.230 & -0.230 & -0.230 \\ 
G6r, G6rcp & 2.27 & 3.10 & 1.15 & 3.36 & 0.407 & -0.201 & -0.201 & -0.201 \\ 
\hline
\end{tabular} \\
\end{center}
\label{tab3}
\end{table}

\begin{table}[p] 

\renewcommand{\baselinestretch}{0.9}

\caption{Parameters of the $\Lambda$-nucleon interaction
($\Lambda$-N).  Strengths $V$ in MeV and ranges $r$ in fm of the
central (c), spin-spin (ss), tensor (t) and spin-orbit (so) parts are
listed. The central part $V^{(l)}_{c}(r)$ depends on the angular
momentum $l=0,1$. The radial shapes are gaussian (G), exponential (E) 
and Yukawa (Y), respectively. The corresponding interactions reproduce
the scattering lengths of the {\em Nijmegen F} model
\protect\cite{nag79} (labels 1 and 3), the {\em Nijmegen SC} model
\protect\cite{mae89} (labels 5 and 6), the reduced Nijmegen models
(label r) and the modified potentials (label ra and rb) corresponding
to Fra, Frb, SCa and SCb in table \protect\ref{tab3}, see also the
text. The interactions with labels 3 and 5 all include d-wave coupling 
in the analysis. Label c refer to interactions with a fictitious repulsive 
term in the central (l=0) part with parameters $V_{cr} = 550$ MeV 
and $r_{cr}= 0.53$ fm. Label p refers to interactions 
reproducing the same sets of low-energy data (including p-wave 
scattering lengths) as the interactions labeled G6 and G6r, 
respectively (see table \protect\ref{tab3}).}

\renewcommand{\baselinestretch}{1.5}

\vspace{5mm} \hspace*{-1.5cm}
\begin{tabular}{c|cccccccccc} \hline
$\Lambda$-N & $V^{(l=0)}_{c}$ & $r^{(l=0)}_{c}$ & $V^{(l=1)}_{c}$ & $r^{(l=1)}_{c}$ & $V_{ss}$ & $r_{ss}$ & $V_{t}$ & $r_{t}$ & $V_{so}$ & $r_{so}$  \\ 
\hline \hline
    G1     & -24.354 & 1.4986 & -5.413  & 1.4986 & 1.931 & 1.5998 & 1.884 &  1.4986 & -2.347 & 1.4986  \\
    Y1     & -50.833 & 0.8726 & -10.010 & 0.8726 & 2.529 & 1.0764 & 3.296 & 0.8726 & -4.109 & 0.8726 \\
    G3     & -24.272 & 1.4972 & -24.272 & 1.4972 & 2.210 & 1.5486 & 2.053 & 1.7278 & -3.644 & 2.0488 \\  
    G1r    & -21.919 & 1.4986 & -5.413 & 1.4986 & 1.931 & 1.5998 & 1.884 & 1.4986 & -2.347 & 1.4986  \\
    E1r    & -86.859 & 0.5678 & -28.426 & 0.5710 & 6.199 & 0.6503 & 7.696 & 0.5710 & -14.828 & 0.5710 \\
    Y1r    & -44.093 & 0.8912 & -10.010 & 0.8726 & 1.997 & 1.2141 & 3.296 & 0.8726 & -4.109 & 0.8726 \\
\hline \hline
    G5     & -28.280 & 1.361  & -28.280 & 1.361  & 4.838 & 1.8221 & 4.364 & 1.4401 & -8.445 & 1.5902 \\ 
    E5     & -97.431 & 0.5446 & -97.431 & 0.5446 & 25.420 & 0.6049 & 15.604 & 0.5740 & -4.964 & 1.0270 \\
    Y5     & -56.291 & 0.8119 & -56.291 & 0.8119 & 4.684 & 1.4193 & 2.347 & 1.1822 & -2.214 & 1.7113 \\ 
    G5r    & -25.452 & 1.361  & -25.452 & 1.361 & 4.838 & 1.8221 & 4.364 & 1.4401 & -8.445 & 1.5902 \\ 
    E5r   & -98.351 & 0.5218 & -98.351 & 0.5218 & 18.896 & 0.6812 & 16.122 & 0.5532 & -2.565 & 1.2230 \\ 
    Y5r   & -50.654 & 0.8145 & -50.654 & 0.8145 & 9.725  & 1.0234 & 0.922 & 1.5597 & -2.733 & 1.6423 \\ 
    G5ra   & -25.185 & 1.3759 & -25.185 & 1.3759 & 5.122 & 1.8748 & 3.745 & 1.4837 & -9.056 & 1.5667 \\ 
    E5ra  & -98.693 & 0.5236 & -98.693 & 0.5236 & 19.195 & 0.7137 & 16.780 & 0.5465 & -2.302 & 1.2634 \\ 
    Y5ra  & -53.974 & 0.7977 & -53.974 & 0.7977 & 8.172  & 1.1573 & 0.318 & 2.2190 & -5.410 & 1.3073 \\ 
    G5rb   & -25.506 & 1.3388  & -25.506 & 1.3388 & 5.443 & 1.8284 & 4.618 & 1.4062 & -8.935 & 1.5383 \\ 
    E5rb  & -96.507 & 0.5172 & -96.507 & 0.5172 & 22.100 & 0.6752 & 17.087 & 0.5391 & -2.390 & 1.2215 \\ 
    Y5rb  & -51.068 & 0.8006 & -51.068 & 0.8006 & 10.151 & 1.0598 & 0.444 & 1.9396 & -5.425 & 1.3061 \\ 
\hline
 G6    & -28.644  & 1.3588 & -28.644  & 1.3588 & 4.694 & 1.8178 & 0 & - & 0 & - \\
 G6c   & -152.876 & 0.9853 & -152.876 & 0.9853 & 5.284 & 1.7492 & 0 & - & 0 & - \\ 
 G6cp  & -152.876 & 0.9853 & -236.971 & 0.8468 & 5.284 & 1.7492 & 0 & - & 0 & - \\ 
\hline
 G6r   & -25.723  & 1.3614 & -25.723  & 1.3614 & 4.574 & 1.8368 & 0 & - & 0 & - \\
 G6rc  & -151.130 & 0.9729 & -151.130 & 0.9729 & 5.094 & 1.7750 & 0 & - & 0 & - \\
 G6rcp & -151.130 & 0.9729 & -319.318 & 0.7722 & 5.094 & 1.7750 & 0 & - & 0 & - \\
\hline
 \end{tabular} \\
\label{tab4}
\end{table}

\begin{table}[p] 
\renewcommand{\baselinestretch}{0.9}
\caption{The contributing three-body channel quantum numbers for the
hypertriton ground state. The upper part of the table refers to the
Jacobi coordinates with the $\Lambda$-particle as spectator, while the
lower part refers to the other two Jacobi systems where one of the
nucleons is the spectator. The quantity $t_x$ is the total isospin of
the two-particle system connected by the relative coordinate
$\protect\mbox{\bf x}$.}
\renewcommand{\baselinestretch}{1.5}
\vspace{5mm}
\begin{center}
 \begin{tabular}{cccccccc} \hline
Channel & Config. & $l_{x}$ & $l_{y}$ & $L$ & $s_{x}$ & $S$ & $t_{x}$  \\ 
\hline
1       & $^{3}S_{1}$ & 0 & 0 & 0 & 1 & 1/2 & 0 \\
2       & $^{1}P_{1}$ & 1 & 1 & 0 & 0 & 1/2 & 0 \\
3       & $^{1}P_{1}$ & 1 & 1 & 1 & 0 & 1/2 & 0 \\
4       & $^{3}D_{1,2}$ & 2 & 0 & 2 & 1 & 3/2 & 0 \\ \hline
5    & $^{1}S_{0}$ & 0 & 0 & 0 & 0 & 1/2 & 1/2 \\
6    & $^{3}S_{1}$ & 0 & 0 & 0 & 1 & 1/2 & 1/2 \\
7    & $^{1}P_{1}$ & 1 & 1 & 0 & 0 & 1/2 & 1/2 \\
8    & $^{1}P_{1}$ & 1 & 1 & 1 & 0 & 1/2 & 1/2 \\
9    & $^{3}P_{j}$ & 1 & 1 & 0 & 1 & 1/2 & 1/2 \\
10   & $^{3}P_{j}$ & 1 & 1 & 1 & 1 & 1/2 & 1/2 \\
11   & $^{3}P_{j}$ & 1 & 1 & 1 & 1 & 3/2 & 1/2 \\
12   & $^{3}P_{j}$ & 1 & 1 & 2 & 1 & 3/2 & 1/2 \\ 
13   & $^{3}S_{1}$ & 0 & 2 & 2 & 1 & 3/2 & 1/2 \\ 
14   & $^{3}D_{1,2}$ & 2 & 0 & 2 & 1 & 3/2 & 1/2 \\ 
\hline
 \end{tabular} \\
\end{center}
\label{tab5}
\end{table}

\begin{table}[p]
\renewcommand{\baselinestretch}{0.85}
\caption{The $^{3}_{\Lambda}H$ binding energy in MeV relative to the 
deuteron for different two-body interactions. 
The $\Lambda$-N interactions of the upper part (double separation line) 
reproduce the low-energy data of the F and SC Nijmegen models, while 
those of the lower part refer to the reduced ones (Fr and SCr), see tables 
\protect\ref{tab1}-\protect\ref{tab4}.
Labels $\alpha$ and $\beta$ indicate which channels of table 
\protect\ref{tab5} are included: "all except 13 and 14", and 
"only 1,4,5,6", respectively. The $\beta$ combination excludes all 
p- and d-states in the $\Lambda$-N subsystem. No greek labels means 
that all channels are included. The $^{3}_{\Lambda}H$ and the 
$\Lambda$-deuteron rms radii, and the $\Lambda$-d $^{2}S$- 
scattering length (negative in our notation) and effective range 
are also showed. All lengths are in fm.}
\renewcommand{\baselinestretch}{1.5}
\vspace{-4mm}
\begin{center}
 \begin{tabular}{cl|ccccc} \hline
\multicolumn{1}{c}{N-N} & \multicolumn{1}{l}{$\Lambda$-N} &
\multicolumn{1}{|c}{$B_{\Lambda}$} & $<\!r^2\!>^{1/2}$ & $<\!r_{\Lambda d}^2\!>^{1/2}$ &
$a_{\Lambda-d}$($^{2}S$) & $r_{\Lambda-d}$($^{2}S$) 
\\ \hline
 G3 &  G1   & 0.31   &  4.03  & 7.72 & -11.5 & 3.2 \\
 G3 &  Y1   & 0.53   &  3.06  & 5.52 & -9.0 & 2.8 \\
 Y3 &  G1   & 0.39   &  3.96  & 7.48 & -10.8 & 3.4 \\
 Y3 &  Y1   & 0.65   &  3.21  & 5.75 & -8.9 & 3.2 \\ 
 GC1 &  G3  & 0.22   &  4.53  & 8.79 & -13.0 & 3.3 \\
 EC1 &  G3  & 0.23   &  4.36  & 8.41 & -12.5 & 3.2 \\
\hline
 GC1 &  G5  & 0.37   &  3.71  & 6.95 & -10.5 & 3.1 \\
 GC1 &  E5  & 0.44   &  3.44  & 6.35 & -9.7 & 2.9 \\
 GC1 &  Y5  & 0.51   &  3.26  & 5.92 & -9.2 & 2.8 \\
 EC1 &  G5  & 0.40   &  3.58  & 6.67 & -10.2 & 3.0 \\
\hline \hline
 G1 & G1r & 0.064 & 7.29 & 14.76 & -22.6 & 3.1 \\
 G1 & G1r $^{(\alpha)}$ & 0.063 & 7.32 & 14.81 & -22.7 & 3.1 \\
 G1 & G1r $^{(\beta)}$ & 0.055 & 7.62 & 15.44 & -24.1 & 3.1 \\
 G1 & Y1r & 0.158 & 5.15 & 10.19 & -14.9 & 3.0 \\
 G1 & Y1r $^{(\alpha)}$ & 0.157 & 5.17 & 10.23 & -14.9 & 3.0 \\
 G1 & Y1r $^{(\beta)}$ & 0.117 & 5.40 & 10.72 & -15.2 & 3.0 \\
 E1 & E1r & 0.114 & 5.86 & 11.72 & -17.2 & 3.0 \\
 E1 & E1r $^{(\alpha)}$ & 0.113 & 5.89 & 11.77 & -17.3 & 3.0 \\
 E1 & E1r $^{(\beta)}$ & 0.101 & 6.14 & 12.32 & -18.2 & 3.0 \\
 G2 & G1r & 0.046 & 8.03 & 16.28 & -26.1 & 3.3 \\
 E2 & G1r & 0.049 & 7.87 & 15.95 & -25.4 & 3.2 \\
 G3 & G1r & 0.091 & 6.55 & 13.14 & -19.3 & 4.8 \\
 G3 & E1r & 0.145 & 5.37 & 10.66 & -15.7 & 4.6 \\
 G3 & Y1r & 0.172 & 5.01 & 9.89 & -15.6 & 4.9 \\
\hline
 GC1  & G5r & 0.108 & 5.88 & 11.73 & -17.8 & 3.3 \\
 GC1  & E5r & 0.161 & 5.09 & 10.01 & -14.9 & 3.2 \\
 GC1  & Y5r & 0.174 & 4.95 & 9.71  & -14.4 & 3.2 \\
 GC1 & G5ra & 0.166 & 5.06 & 9.96 & -14.8 & 3.3 \\
 GC1 & E5ra & 0.235 & 4.41 & 8.53 & -12.7 & 3.1 \\
 GC1 & Y5ra & 0.247 & 4.33 & 8.37 & -12.5 & 3.2 \\
 GC1 & G5rb & 0.084 & 6.25 & 12.52 & -19.3 & 3.3 \\
 GC1 & E5rb & 0.130 & 5.48 & 10.87 & -16.3 & 3.2 \\
 GC1 & Y5rb & 0.149 & 5.23 & 10.32 & -15.3 & 3.2 \\
 EC1 & G5ra & 0.176 & 4.87 & 9.53  & -14.1 & 3.3 \\
 EC1 & G5rb & 0.100 & 5.96 & 11.89 & -18.1 & 3.2 \\ 
\hline
 \end{tabular} \\
\end{center}
\label{tab6}
\end{table}

\begin{table}[p]
\renewcommand{\baselinestretch}{0.9}
\caption{The hypertriton binding energy in MeV relative to the deuteron for
different two-body interactions. The N-N interaction is always the one 
labeled GC1, see table \protect\ref{tab1}. The $\Lambda$-N interactions
all reproduce the s-wave low-energy data of the Nijmegen models SC
and SCr, see table \protect\ref{tab4}. The interactions labeled G6cp 
and G6rcp reproduce the same s-wave low-energy data and p-wave scattering
lengths as the interactions labeled G6 and G6r. The values with the (a) 
include the energy gain obtained adjusting the p-wave scattering lengths 
to resemble the result for the case without a repulsive core.
The first column indicates the partial waves included in the computations.}
\renewcommand{\baselinestretch}{1.5}
\begin{center}
 \begin{tabular}{c|cc|cc|cc} \hline
partial waves & $\Lambda$-N & $B_{\Lambda}$ & $\Lambda$-N & $B_{\Lambda}$ 
& $\Lambda$-N & $B_{\Lambda}$ \\ \hline
 s         & G6  & 0.442 & G6c  & 0.303 & G6cp & 0.303 \\
 s,p       & G6  & 0.484 & G6c  & 0.365 & G6cp & 0.387 \\
 s,p,d     & G6  & 0.490 & G6c  & 0.396 & G6cp & 0.418$^{a}$ \\
 s,p,d,f   & G6  & 0.491 & G6c  & 0.397 & G6cp & 0.419$^{a}$ \\
 s,p,d,f,g & G6  & 0.491 & G6c  & 0.397 & G6cp & 0.419$^{a}$ \\ \hline
 s         & G6r & 0.137 & G6rc & 0.083 & G6rcp & 0.083 \\
 s,p       & G6r & 0.166 & G6rc & 0.119 & G6rcp & 0.136 \\
 s,p,d     & G6r & 0.170 & G6rc & 0.125 & G6rcp & 0.152$^{a}$ \\
 s,p,d,f   & G6r & 0.170 & G6rc & 0.126 & G6rcp & 0.153$^{a}$ \\
 s,p,d,f,g & G6r & 0.170 & G6rc & 0.126 & G6rcp & 0.153$^{a}$ \\
 \hline
 \end{tabular} \\
\end{center}
\label{tab7}
\end{table}

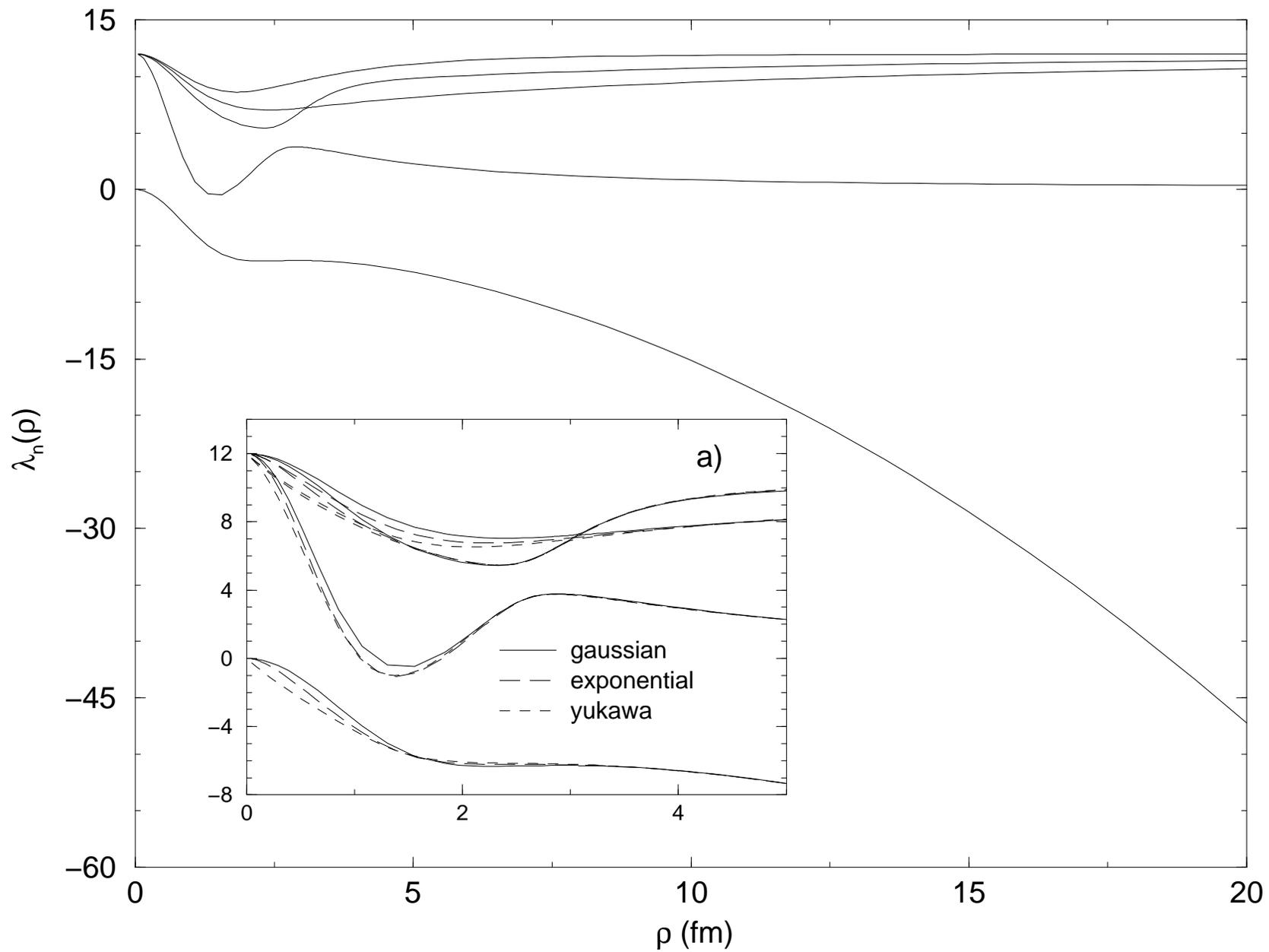
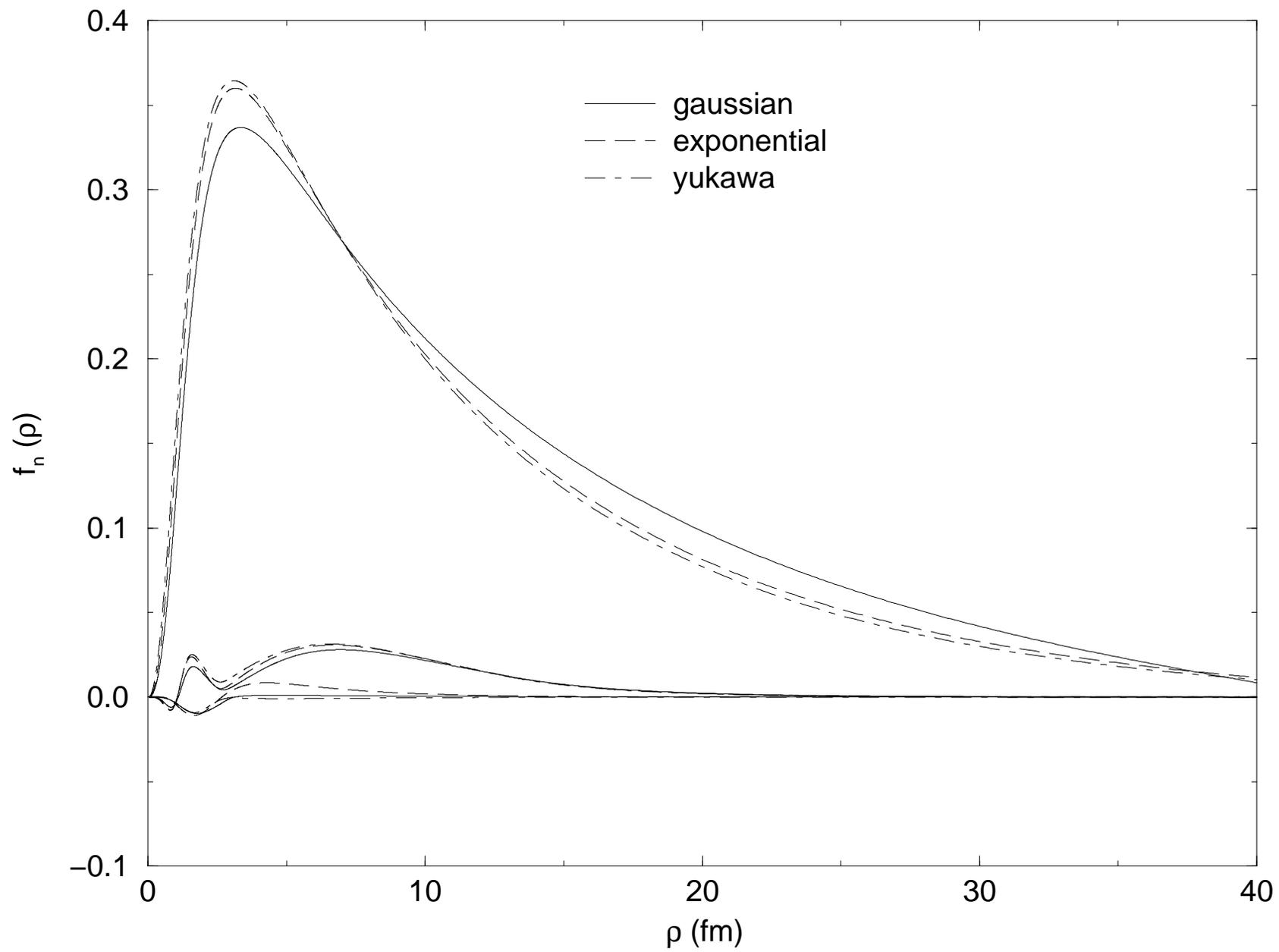
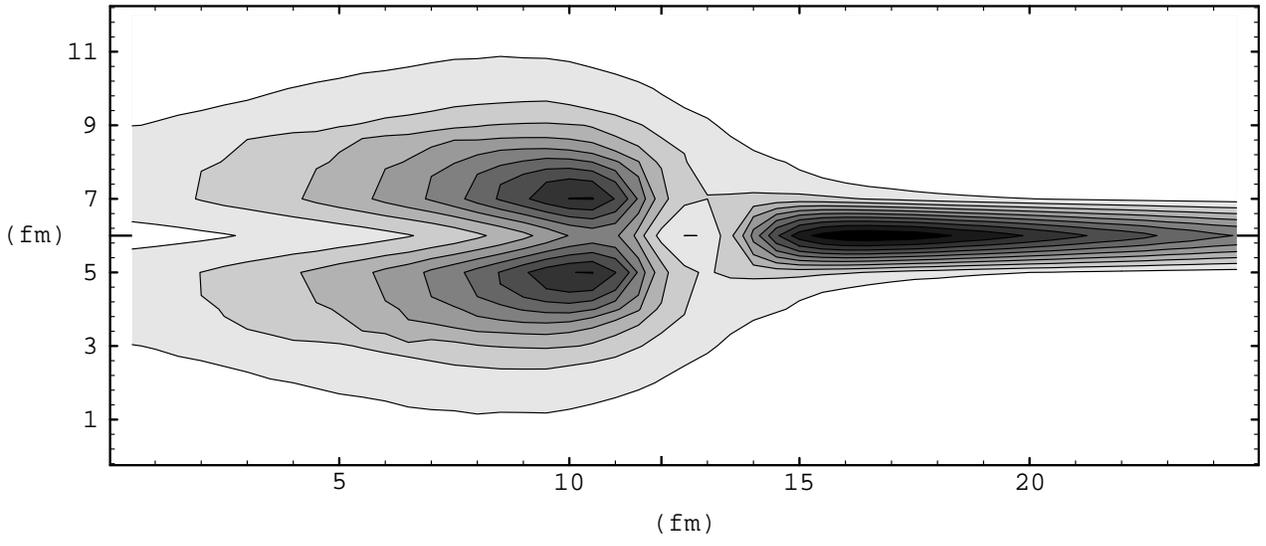
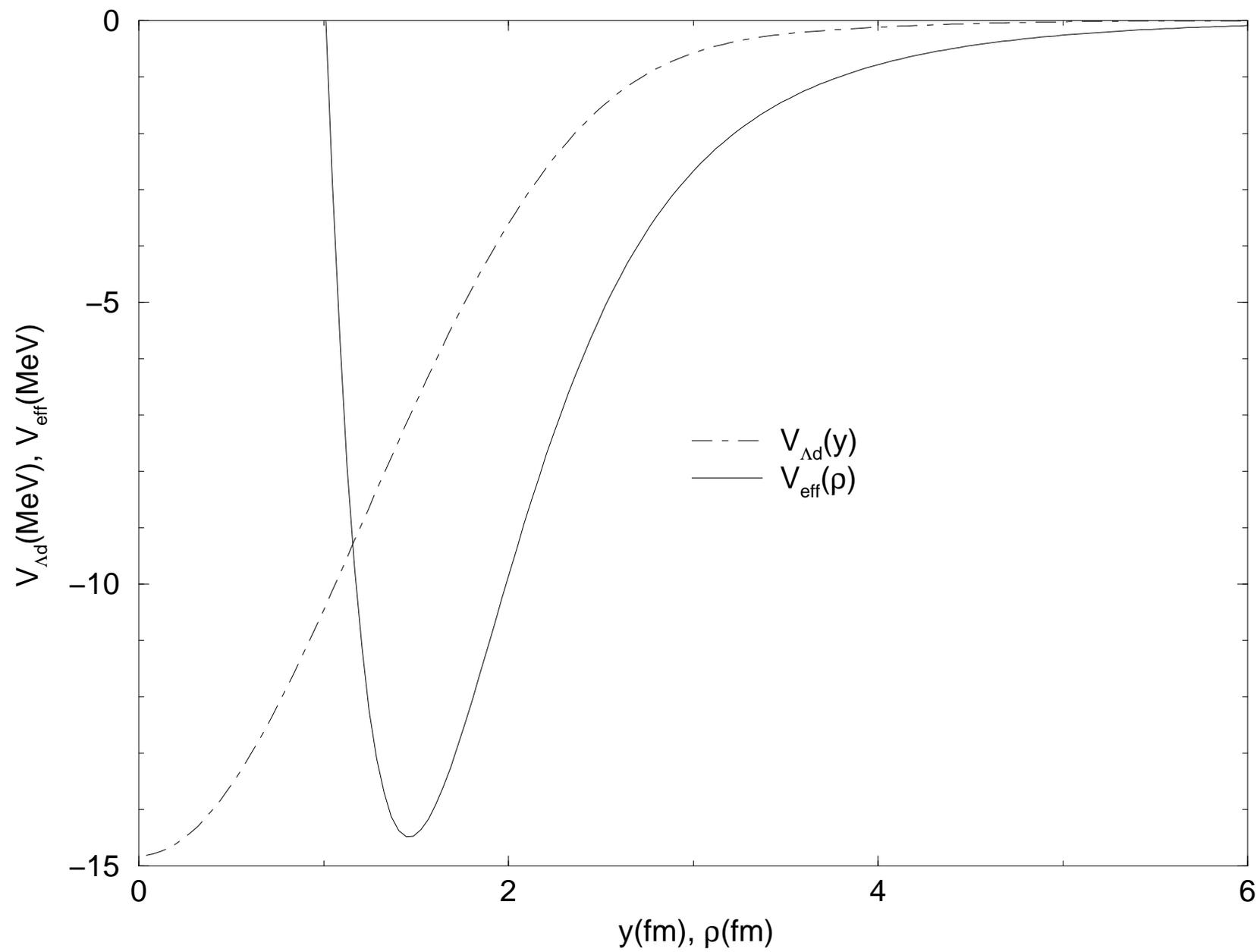
\end{document}